\documentclass{jetpl}
\usepackage{bm}
\usepackage[cp1251]{inputenc}
\usepackage[english]{babel}

\usepackage{graphicx}

\onecolumn

\title{Optical orientation of nuclei in nitrogen alloys GaAsN at room
temperature}

\rtitle{Optical orientation of nuclei in GaAsN at room
temperature}

\sodtitle{Optical orientation of nuclei in nitrogen alloys GaAsN
at room temperature}

\author{V.\,K.\,Kalevich$^{*,\dag,}$\/\thanks{kalevich@solid.ioffe.ru},
        M.\,M.\,Afanasiev$^{*,\dag}$, A.\,Yu.\,Shiryaev$^{*}$, and A.\,Yu.\,Egorov$^{*}$ }

\rauthor{V.\,K.\,Kalevich, M.\,M.\,Afanasiev, A.\,Yu.\,Shiryaev,
A.\,Yu.\,Egorov }

\sodauthor{Kalevich, Afanasiev, Shiryaev, Egorov }

\address{$^{*}$A.\,F.\,Ioffe Physico-Technical Institute, St. Petersburg 194021, Russia}
\address{$^{\dag}$Spin Optics Laboratory, St. Petersburg State University, St. Petersburg 198504, Russia}

\abstract{The intensity and the giant circular polarization of
edge luminescence in a longitudinal magnetic field have been
measured in nitrogen alloys GaAsN under circularly polarized
pumping. It has been found that these dependences are shifted with
respect to zero field by a value $B_{\textrm{eff}}$. The magnitude
of the internal field $B_{\textrm{eff}}$ increases with increase
in pumping intensity and reaches saturation ($\approx250$\,Gauss)
at great densities of excitation. The saturation of the
$B_{\textrm{eff}}$ field with growth of pumping indicates that
this is a field of nuclei, polarized dynamically due to hyperfine
interaction with optically oriented deep paramagnetic centers,
rather than a field of exchange interaction created on the center
by spin-polarized photo-excited conduction electrons. The short
time of nuclear polarization by electrons ($<15\,\mu$s), measured
under modulation of circular polarization of the exciting light
with high frequency, points to a small number of nuclei undergoing
hyperfine interaction with an electron localized at a center. }


\begin{document}

\maketitle

In the last few years the spin properties of nitrogen alloys
Ga(In)AsN have been drawing heightened attention, owing to the
fact that in such alloys there occurs at room temperature an
anomalously great (up to 90\%) spin polarization of free
electrons, which is retained for a long time under optical pumping
(see review \cite{bib:JPCM2010} and references therein).

Recently we have found \cite{bib:PRB2012} that the circular
polarization degree $\rho$ and the intensity $J$ of the edge
photoluminescence (PL) excited by circularly polarized light in
GaAs$_{1-x}$N$_{x}$ crystals ($x\sim1$\%) at room temperature,
increase substantially in a longitudinal magnetic field
$B\sim1$\,kG (for an example see Fig.\,1). This increase depends
on intensity of pumping and can reach a twofold value under weak
or moderate pumping. We considered in \cite{bib:PRB2012} the
suppression of spin relaxation of deep paramagnetic centers in a
longitudinal magnetic field as a possible cause of such a rise in
$\rho$ and $J$. These centers appear in the process of crystal
growing, when nitrogen atoms are introduced into gallium arsenide,
and create a dominant channel for recombination of free carriers.
The recombination is spin-dependent, since its rate is governed by
spin polarization of the electrons localized on centers. In other
words, the paramagnetic centers act as a spin filter, blocking the
capture from the conduction band of electrons with predominant
orientation of spins. Therefore an increase in spin relaxation
time of the centers in the magnetic field brings about a strong
increase in their polarization $P_c$, which, in its turn, is
accompanied by the growth in the spin polarization $P$ of free
electrons and also in their concentration. The former reveals
itself in the rise of $\rho$ (since $\rho\propto P$), and the
latter, in the rise of $J$. We supposed in \cite{bib:PRB2012} that
the spin relaxation of deep centers of GaAsN in the zero field is
due to chaotic magnetic fields, which are generated by spin
fluctuations of the nuclei located in the vicinity of a
paramagnetic center and coupled with the center by the hyperfine
interaction \cite{bib:DP1973,bib:Merk2002}.

Also in \cite{bib:PRB2012} it has been found that the experimental
dependences $J(B)$ and $\rho(B)$ are shifted with respect to zero
field by a value of $|B_{\textrm{eff}}|$$\,\sim\,$100\,Gauss;
besides, the direction of the shifting reverses with reverse of
the sign of circular polarization of exciting light (see Fig.\,1).
It is pointed out in \cite{bib:PRB2012} that such a shift is
possibly caused by the Overhauser stationary field, $B_N$, which
acts on a localized electron from the optically-oriented
crystal-lattice nuclei located near center. The field $B_N\propto
\langle I\rangle$ \cite{bib:DP1973,bib:OO,bib:OO2}, where $\langle
I\rangle$ is the mean nuclear spin, and is directed along the
external field $B$ (along the exciting beam). As a result, adding
up with the external field, or subtracting from it, the field
$B_N$ brings about the shift of the curves $J(B)$ and $\rho(B)$.
As the polarization of nuclei is proportional to the polarization
of electrons, the sign reverse of the pumping polarization
involves a change in the direction of the field $B_N$ and,
consequently, in the direction of the shift of the $\rho(B)$ and
$J(B)$ curves. At the same time, the shift of the $\rho(B)$ and
$J(B)$ curves can also be induced by the exchange interaction of a
polarized paramagnetic center with other paramagnetic centers or
with spin-polarized free electrons in the conduction band. Indeed,
each of these exchange interactions creates at the center an
effective magnetic field $B_{\textrm{ex}}$, which is proportional
to the full spin of localized or free electrons
\cite{bib:AB,bib:Kirill} and hence must change the direction when
the sign of circular polarization of excitation reverses. The
exchange interaction between deep centers in the studied GaAsN
crystals can be neglected on account of the small concentration of
centers ($N_c \approx10^{15}\textrm{cm}^{-3}$ \cite{bib:JPCM2010})
and, accordingly, the small overlap of their wave functions
\cite{bib:AB,bib:Kirill}. One cannot neglect the exchange
interaction with conduction electrons beforehand, since in GaAsN
the exchange interaction constant is unknown \cite{bib:Kir}, while
the experimental observation of the field $B_{\textrm{eff}}$
requires the use of a strong, $\sim$\,50\,mW, pumping. Under such
a pumping the polarization of free electrons approximates 100\%
\cite{bib:JPCM2010}, and their concentration, according to our
evaluation, amounts to a considerable value of
$n\sim10^{15}$\,$\textrm{cm}^{-3}$, which can bring about a great
value of the exchange field.

\begin{figure}
    \centering
    \includegraphics[width=0.4\linewidth]{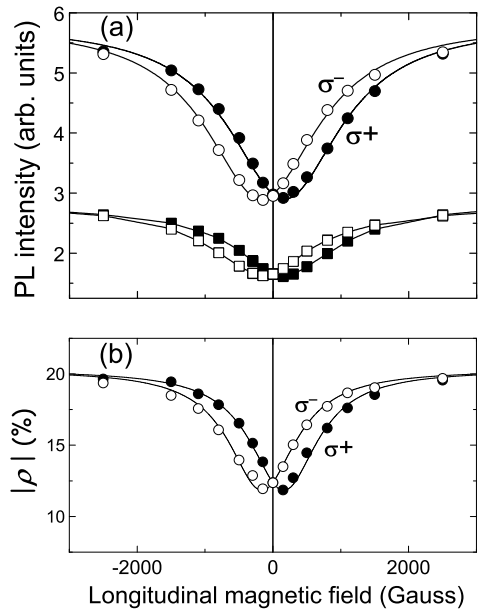}
    \caption{\label{IRo(B)} Fig.1: (a) PL intensity  and (b) PL circular polarization degree
    as a function of a longitudinal magnetic field measured at room temperature in GaAs$_{0.979}$N$_{0.021}$
    under excitation by right-circularly, $\sigma^{+}$ (solid circles) and left-circularly, $\sigma^{-}$
    (open circles) polarized light of the intensity
    $W=30$\,mW. Solid and open squares in Fig.1a present PL intensities detected with the excitation circular polarization
    alternating at the frequency of 35\,kHz
    (see text for details). Excitation energy $h\nu_{\rm exc}=1.393$\,eV (excitation below the GaAs barrier),
detection energy $h\nu_{\rm det}=1.159$\,eV (detection near the PL
band maximum). }
\end{figure}

The present work describes an experiment unambiguously indicative
of the nuclear nature of the field $B_{\textrm{eff}}$. On the
qualitative level, we can discriminate between the nuclear field
$B_{N}$ and the field of exchange interaction with free electrons
$B_{ex}$ by their dependence on pumping intensity. Indeed, the
hyperfine contact ($A\textbf{s}_c$\,$\cdot$\,$\textbf{I}$) and
exchange ($Q\textbf{s}_c$\,$\cdot$\,$\textbf{s}$) interactions
create the fields $B_{N}\propto\langle I\rangle \propto P_c$ and
$B_{ex}\propto S =Pn/2$ \cite{bib:AB}, which act upon the spin
$\textbf{s}_c$ of an electron localized on the center. Here
$\textbf{I}$ and $\textbf{s}$ are the spins of a nucleus and a
free electron, $A$ and $Q$ are the hyperfine and exchange
constants, $S$ is the total spin of free electrons, $P_c$ and $P$
is the polarization of localized and free electrons, respectively.
Under sufficiently large pump intensity $W$ (in our case at
$W\gtrsim75$\,mW) the spin filter effect in GaAsN leads to
practically full polarization of both localized ($P_c\simeq1$) and
free ($P\simeq1$) electrons \cite{bib:JPCM2010}. With further
increase of pumping the field $B_{N}$ has to remain unaltered,
while the field $B_{ex}$ has to grow linearly as the excitation
intensity increases, because $n\propto W$: $B_{N}=\textrm{const}$,
$B_{ex}\propto W$. This serves as a main criterion for separation
of the nuclear and exchange fields.

We have investigated an undoped alloy GaAsN with a nitrogen
content of 2.1\%, grown as a film 0.1\,$\mu$m thick on a
semi-insulating GaAs substrate \cite{bib:Egorov2005}. Polarization
of free electrons $P$ was induced by interband absorption of
circularly polarized light \cite{bib:OO,bib:OO2}. Measurements
were taken of the intensity $J$ and the degree of circular
polarization of the edge PL $\rho=P'P$, where the numerical factor
$P'$\,$\leq$\,1 \cite{bib:OO,bib:OO2}. The degree $\rho$ is
defined as $\rho=(J^+ -J^-)/J$, where $J^+$ and $J^-$ are PL
components polarized on the right ($\sigma^+$) and the left
($\sigma^-$) circle, $J=J^+ +J^-$. A continuous-wave (CW)
titanium-sapphire laser was used to excite photoluminescence,
which was registered by means of a photomultiplier with InGaAsP
photocathode. The values of $\rho$ and $J$ were measured using a
high-sensitive polarization analyzer \cite{bib:Kulkov} comprising
a photoelastic polarization modulator \cite{bib:Jasperson} and a
lock-in two-channel photon counter. The measurements were carried
out at 300\,K under perpendicular incidence of the laser beam onto
the sample and PL recorded in opposite direction. The magnetic
field was directed along the exciting beam (Faraday geometry).

\begin{figure}
  \centering
    \includegraphics[width=0.85\linewidth]{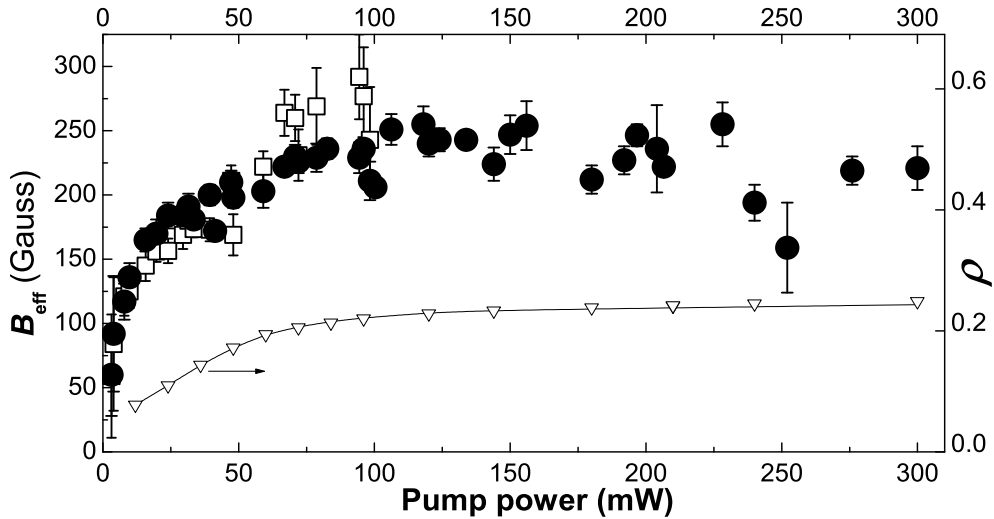}
    \caption{\label{Shift(J)} Fig.2: The shifts of the curves $I(B)$ (solid circles) and $\rho(B)$
(open squares) with respect to zero of the magnetic field in
GaAs$_{0.979}$N$_{0.021}$ at room temperature under different
powers of right-circularly polarized excitation. Triangles show
the dependence $\rho(J)$ taken at $B=0$; solid line is a guide for
the eye. $h\nu_{\rm exc}=1.393$\,eV, $h\nu_{\rm det}=1.159$\,eV. }
\end{figure}

The field magnitude $|B_{\textrm{eff}}|$ was found through fitting
the experimental dependences $J(B)$ and $\rho(B)$ by Lorentzians
of the form
$y(B)=y_{\textrm{max}}+(y_{\textrm{min}}-y_{\textrm{max}})/[1+(B-B_{\textrm{eff}})^2/B_{1/2}^2]$
(solid lines in Fig.\,1), where
$y_{\textrm{min}}=y(B$~=~$B_{\textrm{eff}})$,
$y_{\textrm{max}}=y(B\rightarrow\infty)$ and $B_{1/2}$ is the
half-width of the curve on the half-height. Fig.\,2 shows
dependences $B_{\textrm{eff}}(W)$ which obtained as a result of
fitting the curves $J(B)$ (solid circles) and $\rho(B)$ (empty
squares) measured in GaAs$_{0.979}$N$_{0.021}$ under the  change
in intensity (from 3 to 300\,mW) of right-hand ($\sigma^{+}$)
polarized excitation. We see that $B_{\textrm{eff}}$ tends to zero
with decreasing $W$. This is a result of the decline in efficiency
of spin-dependent recombination under weak pumping, in consequence
of which the polarization of localized electrons $P_c$ tends to
zero \cite{bib:JPCM2010}. With increasing pumping the field
$B_{\textrm{eff}}$ increases and reaches saturation at
$W\approx75$\,mW. The dependence $\rho(W)$ (triangles in Fig.\,2)
measured at $B=0$ is saturated at the same intensity, which speaks
of practically full polarization of localized electrons, $P_c
\simeq1$ \cite{bib:JPCM2010}, at $W\gtrsim75$\,mW. Therefore the
saturation of field $B_{\textrm{eff}}$ at $W\gtrsim75$\,mW permits
us to draw an unambiguous conclusion that a dominating role in
formation of the field $B_{\textrm{eff}}$ belongs to the nuclear
polarization.

An important parameter of the nuclear spin system is the time of
its polarization by optically oriented electrons $T_{1e}$. As a
rule, the process of optical orientation of nuclei under interband
pumping in semiconductors and semiconductor nanostructures is very
slow: according to the concentration of doping impurity and the
crystal temperature, the time of longitudinal nuclear relaxation
may range from seconds to hours \cite{bib:Abragam}. The inertness
of nuclear spin system frequently serves as a means of
obliterating its polarization. To this purpose the sign of
circular polarization of the exciting light should be varied with
high frequency. In that case the nuclear polarization has no time
to follow the quick alternation of electron polarization, and the
nuclear field $B_N$ comes to zero \cite{bib:OO,bib:OO2}. We
changed the sign of circular polarization of the exciting light
with a frequency 35\,kHz, passing a linearly polarized laser beam
through the photoelastic modulator of polarization
\cite{bib:Jasperson} working on that frequency. The two-channel
photon counter \cite{bib:Kulkov} synchronized with the
polarization modulator permitted the PL intensity to be measured
separately during adjacent ($\sigma^+$ or $\sigma^-$) half-periods
of polarization changing, by opening each of the channels for the
same time equal to $0.4T^*$ where $T^*$ is the oscillation period
of the modulator.

The influence of modulation of pumping polarization on the
magnitude of the field $B_{\textrm{eff}}$ can be best followed by
measuring the magnet-field dependence of luminescence intensity
$J(B)$, applying no other polarization elements but the
polarization modulator installed in the excitation channel (the
registration channel is not equipped with a quarter-wave plate and
linear polarizer of light that are indispensable for measurement
of $\rho$). Fig.\,1a shows the dependences of luminescence
intensity $J^+ (B)$ (solid squares) and $J^- (B)$ (open squares),
accumulated, respectively, in $\sigma^+$ and $\sigma^-$
half-periods of modulation at $W=30$\,mW. It is evident that these
dependences are shifted in opposite directions with respect to the
axis of ordinates by the same magnitude equal to $\approx165$\,G.
Within the limit of measurement error this magnitude coincides
with the magnitude of shift $\approx175$\,G measured under the
constant circular polarization of pumping with the same intensity
(see Fig\.2). Thus, the change in the sign of circular
polarization of pumping with a half-period of 15\,$\mu$s is not
accompanied by suppression of nuclear polarization. It means that
the spin relaxation time of nuclei on electrons $T_{1e}$ is
shorter than 15\,$\mu$s. The short time $T_{1e}$ is a fingerprint
of a strong localization of the electron wave function on a center
and, hence, of a small amount of nuclei undergoing the hyperfine
interaction with the localized electron \cite{bib:OO,bib:Abragam}.

The size of the region of electron localization can be evaluated
by the order of magnitude on the assumption that the localized
electrons lose polarization through relaxation on the chaotic
fluctuations of the nuclear field, and the magnitude $B_f$ of
these fluctuations  presets the half-width  $B_{1/2}$ of the
curves $J(B)$ and $\rho(B)$: $B_f$~$\sim$~$ B_{1/2}$. The
characteristic value of the nuclear field fluctuation $B_f \approx
(B_{N\rm{max}}/N)\sqrt{N}=B_{N\rm{max}}/\sqrt{N}$, where $N$ is
the number of nuclei in the localization region, $B_{N\rm{max}}
\propto A/(g_c \mu_ 0)$ is the field of totally polarized nuclei
acting on the spin of localized electron with \emph{g}-factor
$g_c$, $\mu_0$ is the Bohr magneton, $B_{N\rm{max}}/N$ is the
field created by one nucleus, $A\approx100$\,$\mu$eV in GaAs
\cite{bib:OO}. The $B_{N\rm{max}}$ value doesn't depend on the
size of the region of electron localization \cite{bib:OO}. Using
$B_{N\rm{max}}\approx53$\,kG calculated for GaAs in
\cite{bib:Paget} (see also \cite{bib:OO,bib:OO2}), where $\mid
g_c\mid =0.44$, and taking into account that in GaAsN $g_c =2$
\cite{bib:Nature2009}, one can find $B_{N\rm{max}}\approx12$\,kG
for GaAsN. Taking typical experimental value $B_{1/2}
\approx1000$\,G (see Fig.\,1), we obtain that the field $B_f$ of
the same value is created in GaAsN by
$N\approx(B_{N\rm{max}}/B_f)^2\sim100$ nuclei. On the other hand,
$N\sim2V/v_0$, where $V$ is the volume of electron localization,
$v_0$ is the elementary cell volume. Assuming $V\approx4a_0^3$,
where $a_0$ is the electron Bohr radius, we find $a_0 \sim7$\,\AA.

So small an extent of electron localization and, as a consequence,
a small amount of nuclei, necessitates an alternative explanation
to be considered for the above experimental results; this
explanation employs the limiting case of hyperfine interaction of
a localized-on-center electron with only one nucleus, namely, the
nucleus of the center. The Hamiltonian of such an interaction in
the presence of an external magnetic field \textbf{B} has the form
\cite{bib:AB}: $A\textbf{s}_\textbf{c}\textbf{I}+g_\textrm{c}
\mu_{0} \textbf{B}\textbf{s}_\textbf{c}$ (Zeeman energy of the
nucleus is assumed to be negligible as compared to the energy of
its hyperfine interaction with the electron). For such a case, a
theory of optical orientation of electrons and nuclei in a
semiconductor was developed in \cite{bib:DP1972} by Dyakonov and
Perel. It is based on a partial destruction of the optical
orientation of localized electrons in a zero magnetic field due to
mixing of eigenstates of the system one electron -- one nucleus.
The energy of these states at $B=0$ is determined by the total
spin of electron and nucleus $M=1/2+I$, where $I$ is the nuclear
spin, which for an odd $I$ takes two values separated by an
interval $\delta\sim A$ ($\delta=AM$ \cite{bib:AB}. For example,
$\delta=A$ and $2A$ for $I=1/2$ and $3/2$, respectively). The
destruction of electron orientation occurs effectively, if the
characteristic time of mixing $\tau^* =\hbar/\delta$ is much
shorter than the lifetime of localized electrons $\tau_c$. At
$A\approx 100$\,$\mu$eV we obtain $\tau^* \sim 1$\,ps. In GaAsN
the time $\tau_c$ decreases drastically with the excitation
intensity increasing \cite{bib:JPCM2010}; according to our
evaluation, however, it extends $50$\,ps even at the maximum power
of pumping $W=300$\,mW. It means that $\tau^* <<\tau_c$, and the
mixing of states and also the loss of electron spin polarization
may take place in GaAsN effectively at all employed intensities of
excitation. If $\langle S_0 \rangle$ is the mean spin of electrons
in a moment of their capture by the center, then at $\tau^*
<<\tau_c$ the electron spin decreases down to  $\langle
S_0\rangle/2$, $3\langle S_0\rangle/8$ and $\langle S_0\rangle/3$
for $I=1/2, 3/2$ and $I>>1$, respectively \cite{bib:DP1972}.

The mixing of states slows down upon application of a longitudinal
magnetic field, which ruptures hyperfine coupling thus restoring
polarization of electrons \cite{bib:DP1972,bib:Ivch1979}. In this
case the half-width $B_{1/2}$ of the curves $J(B)$ and $\rho(B)$
determines the order of magnitude for the  external magnetic field
in the presence of which the Zeeman energy of an electron becomes
equal to the energy of hyperfine interaction: $g_c \mu_0
B_{1/2}\sim \delta$. The value of $\delta$, evaluated from the
half-width $B_{1/2}\sim 1000$\,G of experimental curves $J(B)$ in
Fig.\,1a, is $10$\,$\mu$eV.

The deep paramagnetic center, responsible for the spin-dependent
recombination in GaAsN, is the self-interstitial defect Ga$^{2+}$
\cite{bib:Nature2009}. The nucleus of this defect is represented
by two isotopes, $^{69}$Ga and $^{71}$Ga, each with spin $I=3/2$.
The analysis of the EPR spectra of that center has revealed the
fact that up to 20\% of an electron is localized on its nucleus
\cite{bib:Nature2009}. Such a density of the electron on the
nucleus corresponds to the energy of hyperfine interaction
equaling $\approx10$\,$\mu$eV (we are taking into account that the
hyperfine interaction constants for Ga isotopes are
$A_{^{69}\rm{Ga}}\approx38$\,$\mu$eV and
$A_{^{71}\rm{Ga}}\approx49$\,$\mu$eV \cite{bib:Paget}), which
coincides in the order of magnitude with the value of $\delta$,
evaluated above from the half-width $B_{1/2}$ of the curve $J(B)$.
This indicates that the hyperfine interaction of an electron with
one nucleus enables us to describe qualitatively the growth of
experimental relations $J(B)$ and $\rho(B)$.

The Dyakonov-Perel theory \cite{bib:DP1972} predicted also that
the decay of optical orientation of electrons in zero field,
involved by the mixture of states, is accompanied by the
appearance of stationary polarization of nuclear spins in an
ensemble of centers. In the simplest case $I=1/2$ and
$\tau_c/\tau^*>>1$ the mean nuclear spin $\langle I \rangle
=\langle S_0 \rangle \nu_{1e}/(\nu_{1e}+\nu_{1})$
~\cite{bib:DP1972}, where $\nu_{1e}=f/2$ is the decay rate of
nuclear orientation due to hyperfine interaction with captured
nonoriented electrons, $\nu_{1}$ is the rate of nuclear relaxation
in the absence of illumination, $f$ is the number of electrons
captured on the center per unit time. Since the capture frequency
$f$ and, hence, $\nu_{1e}$ are proportional to the light intensity
$W$, $\langle I \rangle$ grows linearly at low intensities and
reaches saturation at a level $\langle S_0 \rangle$ under strong
light when $\nu_{1e}>>\nu_{1}$.

As the appearance of stationary nuclear polarization is brought
about by the mixing of states, the suppression of mixing in a
longitudinal magnetic field $B$ reduces the value of $\langle I
\rangle$ \cite{bib:DP1972}. This is in line with the decrease in
relaxation rate $\nu_{1e}$, which, according to \cite{bib:DP1972},
takes form $\nu_{1e}=(f/2)[1/(1+(g_c \mu_0 B/A)^2]$ for $I=1/2$
and $\tau_c/\tau^*>>1$. In particular, it means that in a
longitudinal field a greater intensity of excitation will be
required to reach the same polarization of nuclei found at $B=0$.
The expression for $\langle I \rangle$ at $I>1/2$ is nowhere to be
found in literature.

The theory \cite{bib:DP1972} infers that the nuclear polarization
may cause asymmetry in the dependence of electron polarization on
the longitudinal magnetic field if the nucleus of a center alone
is taken into account. Such dependence, however, has not been
presented explicitely. In the case of a great amount of nuclei in
the region of localization ($N>>1$) the influence of nuclear
stationary polarization on the orientation of electrons shows up
through the effective magnetic field of nuclei $B_N$ which adds up
to the external field $B$ \cite{bib:DP1973} (see also
\cite{bib:OO,bib:OO2}). At $N=1$ for $\langle I\rangle \neq0$
there exists also an electron spin splitting averaged over nuclear
spin projections, which can be interpreted as the effect of the
mean nuclear field $B_N\propto\langle I\rangle$. As with $N>>1$,
the field $B_N$ attenuates or strengthens the effect of an
external magnetic field upon the electron spin and, as a
consequence, shifts  the curves $J(B)$ and $\rho(B)$ relative to
zero of the field. It has been noted above that at $N=1$ the
nuclear polarization $\langle I\rangle$ in the field $B=0$
increases linearly with increasing pumping under a weak pumping
and gets saturated under a strong pumping. Therefore the magnitude
of the shift must also arrive at saturation with growing $W$. Thus
the model of hyperfine interaction with the nucleus of the center
can explain not only the increase but also the shift of
dependences $J(B)$ and $\rho(B)$, that we have discovered in
GaAsN.

An additional investigation is required in order to elucidate the
relative contributions of the center nucleus and the nearest-to-it
nuclei of the crystal lattice into the optical orientation of a
localized electron and its asymmetry in the longitudinal field in
GaAsN. The observation of forbidden electron magnetic transitions
in weak magnetic fields $B<B_{1/2}$, i.e., in the fields of the
order of some hundreds or tens Gauss (e.g. by means of
electrically \cite{bib:Vlasenko} or optically \cite{bib:Awsch}
detectable EPR) could give a direct corroboration of the mixing
effect.

In the case of $N=1$ the nuclear polarization, being maximal in
the zero field, decays at $B>>B_{1/2}$ \cite{bib:DP1972}, while
the reverse situation is observed for $N>>1$: here the nuclear
polarization is at its maximum in a strong magnetic field
\cite{bib:DP1973,bib:OO,bib:OO2}. Measurement of the $\langle
I\rangle$ magnitude in the weak and the strong field $B$ could
also give arguments in favor of one or the other scenario.

The initial polarization of electrons $\langle S_0 \rangle$ in
both models of optical orientation presented in \cite{bib:DP1973}
and \cite{bib:DP1972} is determined only by selection rules under
interband absorption. In GaAsN the polarization of an electron
captured on the center changes drastically due to the
spin-dependent recombination \cite{bib:JPCM2010}. Therefore, in
order to describe quantitatively the experimental results we have
obtained in GaAsN, it is necessary to modify the model of optical
orientation of electrons in the presence of spin-dependent
recombination through the deep paramagnetic center
\cite{bib:JPCM2010}. This can be done by way of taking into
account the hyperfine interaction of the localized electron both
with the center nucleus and with the field fluctuations of nuclei
situated in the vicinity of the center.

The authors are grateful to E.\,L.~Ivchenko, K.\,V.~Kavokin and
L.\,S.~Vlasenko for fruitful discussions, M.\,S.~Bazlov for the
help in the treatment of experimental results. This research was
supported by the programs of the Russian Academy of Sciences and
the Russian Foundation for Basic Research.

\vfill\eject


\begin{thebibliography}{9}
\bibitem{bib:JPCM2010}
E.\,L. Ivchenko, V.\,K. Kalevich, A.\,Yu. Shiryaev et al., J.
Phys.: Condens. Matter {\bf 22}, 465804 (2010).

\bibitem{bib:PRB2012}
V.\,K. Kalevich, M.\,M. Afanasiev, A.\,Yu. Shiryaev et al., Phys.
Rev. B {\bf85}, 035205 (2012).

\bibitem{bib:DP1973} M.\,I. Dyakonov and V.\,I. Perel, Zh. Exsp. Teor. Fiz. {\bf 65}, 362
(1973) [Sov. Phys. JETP {\bf 38}, 177 (1974)].

\bibitem{bib:Merk2002} I.\,A. Merkulov, Al.\,L. Efros, and M. Rosen, Phys.
Rev. B {\bf65}, 205309 (2002).

\bibitem{bib:OO}
{\it Optical orientation}, eds. F.\,Meier and B.\,Zakharchenya
(North-Holland, Amsterdam, 1984).

\bibitem{bib:OO2}
\emph{Spin Physics in Semiconductors}, ed. M.\,I. Dyakonov,
Springer Series in Solid-State Science, vol. 157 (Springer,
Berlin, 2008).

\bibitem{bib:AB}
A.\,Abragam and B.\,Bliney, \emph{Electron Paramagnetic Resonance
of Transition Ions} (Clarendon, Oxford, 1970).

\bibitem{bib:Kirill}
K.\,V. Kavokin, Semicond. Sci. Technol. {\bf 23}, 114009 (2008).

\bibitem{bib:Kir} As far as we know, there is no mention in literature of the direct
experimental measurement of this constant in GaAsN. Theoretical
magnitude of the exchange integral depends strongly on the radius
of the localized electron and the wave function structure of the
deep center, which are also unknown exactly. Assuming that the
ground state of the deep center is situated in the middle of the
band gap, (the binding energy $\varepsilon_c \gtrsim 0.5$\,eV) one
can evaluate $B_{ex}\lesssim 10$\,G. As the binding energy
decreases, the magnitude of the field $B_{ex}$ increases. Hence,
not knowing exactly the magnitude of $\varepsilon_c$, we cannot
reject a priori the contribution of exchange interaction. The
authors are grateful to K.\,V. Kavokin who has evaluated the
magnitude of the field $B_{ex}$.

\bibitem{bib:Egorov2005}
A.\,Yu. Egorov, V.\,K. Kalevich, M.\,M. Afanasiev et al., J. Appl. Phys. {\bf
98}, 013539 (2005).

\bibitem{bib:Kulkov} V.\,D. Kul'kov and V.\,K. Kalevich, Prib. Tekh. Eksp. {\bf 5},
196 (1980) [Instrum. Exp. Tech. {\bf 24}, 1265 (1981)].

\bibitem{bib:Jasperson} S.\,N. Jasperson and S.\,F. Schnatterly, Rev. Sci. Instrum.
{\bf 40}, 761 (1969).

\bibitem{bib:Abragam} A.\,Abraham, \emph{The Principles of Nuclear Magnetism} (Clarendon,
Oxford, 1961).

\bibitem{bib:Paget} D.\,Paget, G.\,Lampel, B.\,Sapoval et al.,
Phys. Rev. B, 5780 {\bf 15} (1977).

\bibitem{bib:Nature2009} X.\,J. Wang, I.\,A. Buyanova, F.\,Zhao et al., Nature Materials
{\bf 8}, 198 (2009).

\bibitem{bib:DP1972} M.\,I. Dyakonov, V.\,I. Perel, Zh. Exsp. Teor. Fiz. {\bf 63}, 1883 (1972).

\bibitem{bib:Ivch1979} E.\,L. Ivchenko, E.\,V. Maksimov, and V.\,N. Medvedev, Opt.
Spectrosc. (USSR) {\bf 47}, 609 (1979).

\bibitem{bib:Vlasenko} H.\,Morishita, L.\,S. Vlasenko, H.\,Tanaka et al., Phys.
Rev. B {\bf80}, 205206 (2009).

\bibitem{bib:Awsch} W.\,F. Koehl, B.\,B. Buckley, F.\,J. Heremans et al., Nature {\bf479}, 84 (2011).

\end{thebibliography}
\end{document}